\documentclass[aps,superscriptaddress,superbib,twocolumn,groupedaddress]{revtex4}

\bibpunct{}{}{,}{s}{}{}

\usepackage{amsmath,amssymb}
\usepackage{graphicx}
\graphicspath{{fig/}}

\usepackage{dcolumn}
\usepackage{bm}
\usepackage{units}
\usepackage{journames}

\usepackage[final]{LLNdef}
\usepackage[final]{modifications}

\begin{document}

\renewcommand\topfraction{0.8}
\renewcommand\bottomfraction{0.7}
\renewcommand\floatpagefraction{0.7}

\title{Kinetic self-organization of trenched templates for the fabrication of versatile ferromagnetic nanowires}%

\author{B. Borca}
\author{O. Fruchart}
\email[]{Olivier.Fruchart@grenoble.cnrs.fr}
\author{Ph. David}
\author{A. Rousseau}
\author{C. Meyer}
\affiliation{Institut N{\'e}el, Department of Nanosciences (CNRS-UJF-INPG), BP166, F-38042 Grenoble
Cedex 9, France}

\date{\today}

\begin{abstract}

We have self-organized versatile magnetic nanowires, \ie with variable period and adjustable
magnetic anisotropy energy~(MAE). First, using the kinetic roughening of W(110) uniaxial templates
of trenches were grown on commercial Sapphire wafers. Unlike most templates used for
self-organization, those have a variable period, \thicknm{4-12} are demonstrated here. Fe
deposition then results in the formation of wires in the trenches. The magnitude of MAE could be
engineered up or down by changing the capping- or underlayer, in turn affecting the mean
superparamagnetic temperature, raised to \tempK{175} so far.

\end{abstract}


\maketitle

\vskip 0.5in

\vskip 0.5in


The bottom-up approach is promising for the fabrication of nanostructures at moderate cost, with
better resolution and less microscopic defects than with lithography. In epitaxial
self-organization~(SO) the building blocks are atoms that aggregate to each other during growth, a
process that can be engineered to fabricate wires and dots with lateral dimensions from the micron
size\cite{bib-FRU03c} down to the atomic size\cite{bib-GAM03b,bib-GAM03,bib-REP00}. The interest
of SO for fundamental research is often the narrow size dispersion\cite{bib-WEI05}, so that
\softremove{in the absence of significant interactions between the nanostructures} macroscopic
measurements reveal the properties of one single nanostructure. \softremove{The latter can indeed
often not be probed directly because of a lack of sensitivity or spatial resolution. }Issues like
the increase of orbital momentum and magnetic anisotropy energy~(MAE) at atomic edges or kinks
have been addressed\cite{bib-GAM03b,bib-GAM03,bib-WEI05}. Concerning applications it is sometimes
argued that SO arrays could be used by addressing single nanostructures one by one, to store one
bit of information for example. A more realistic view is the use of the array as a material with
specific properties arising from the nanoscale. This is the case of semiconductor quantum dots
with lasing properties\cite{bib-MAS02}.

Do magnetic SO systems meet the requirements of applied materials? A prerequisite is that some
versatility of geometrical as well as physical properties is achieved, like tuning the magnitude
of the MAE. We focus here on the fabrication and properties of wires, which lie at the background
of the fast-developing field of spin electronics making use of the propagation of domain walls in
wires for storage or logics devices\cite{bib-CRO05,bib-ALL05,bib-PAR04}.

SO magnetic wires are often achieved by step-decoration of vicinal
surfaces\cite{bib-HAU98,bib-DAL00,bib-FRU04}. This approach is not versatile as a new crystal has
to be prepared with a specific miscut whenever the period needs to be changed. \softremove{Ordered
linear patterns with an adjustable period were reported upon adsorption like for
O/Cu(110)\cite{bib-KER91}, however with no convincing transfer to magnetic materials. } Templates
resulting from kinetic effects are potentially more versatile as the period can be changed with
processing parameters, like temperature. This has been explored using ion etching under grazing
incidence to create ripples on surfaces, independently of crystalline directions\cite{bib-MOR03}.
However a significant control of the period has not been demonstrated so far. Here we explore a
new approach, based on the growth of body-centered-cubic (110) materials on nominally-flat Saphir
yielding parallel trenches with an adjustable period, which we then use to grow magnetic
nanowires. We could tune the magnitude of the MAE of the wires using suitably-chosen capping
layers or underlayers. The use of commercial wafers while achieving some control on both the
period and the MAE represents a significant advancement in the versatility of SO nanowires.

The samples were grown in a set of ultra-high vacuum chambers using pulsed-laser deposition with a
Nd-YAG laser~($\lambda=\unit[532]{nm}$). The chambers are equipped with a quartz microbalance,
sample heating and a translating mask for the fabrication of wedge-shaped samples. A 10 keV
Reflection High Energy Electron Diffraction (RHEED) setup with a CCD camera synchronized with
laser shots permits operation during deposition. An Omicron room-temperature Scanning Tunneling
Microscope (STM-1) and  an Auger Electron Spectrometer (AES) are available. The metallic films are
grown on sapphire \textsl{nominally-flat} $(11\overline20)$ commercial wafers. A detailed
description of the chambers and growth procedures can be found in \cite{bib-FRU07}. The magnetic
measurements were performed ex-situ on \unit[5]{nm}-Mo-capped samples, with a Quantum Design
Superconducting QUantum Interference Device (SQUID) magnetometer.

Ordered arrays of wires were obtained in three steps, consisting of the fabrication of 1.~a
non-magnetic flat buffer layer; 2.~a non-magnetic template displaying a uniaxial array of
trenches; 3.~Fe wires by layer-by-layer deposition at the bottom of the trenches.

The first step is the preparation of a smooth buffer layer. A seed layer of Mo (nominal thickness
$\Theta=\unit[1]{nm}$) followed by W ($\Theta\approx\unit[10]{nm}$) are deposited on sapphire at
room temperature (RT) followed by annealing at $800^\circ$C. This yields a $(110)$ surface of
quality similar to that of metal single crystals, with atomically-flat terrasses of width up to
hundreds of nanometers\cite{bib-FRU98b,bib-FRU07}.

\begin{figure}
  \begin{center}
  \includegraphics[width=81mm]{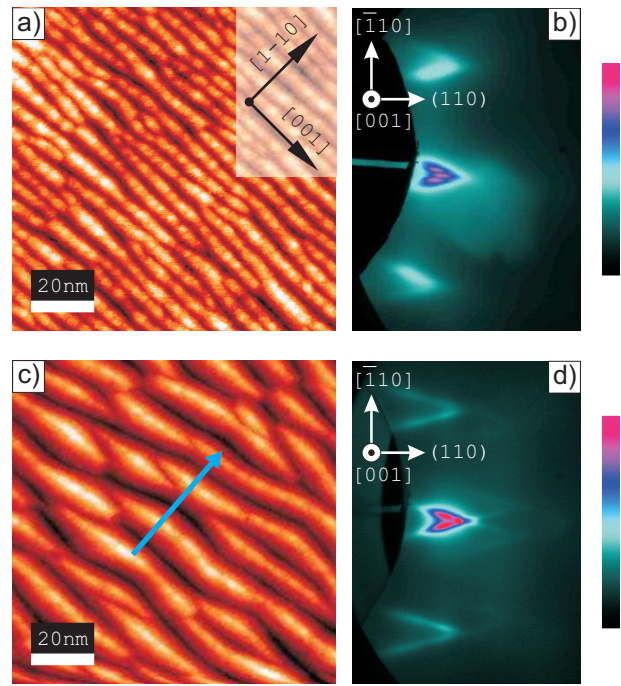}%
  \caption{\label{fig-template}W(110) templates: (a,c) $\thicknm{100\times100}$ STM images of templates with
  small~(a) and large~(c) period, see text (b,d) RHEED patterns along the $[001]$ azimuth for surfaces a and c, resp.
  The sample lies vertical, on the left of the patterns. }
  \end{center}
\end{figure}

\begin{figure}
  \begin{center}
  \includegraphics[width=85mm]{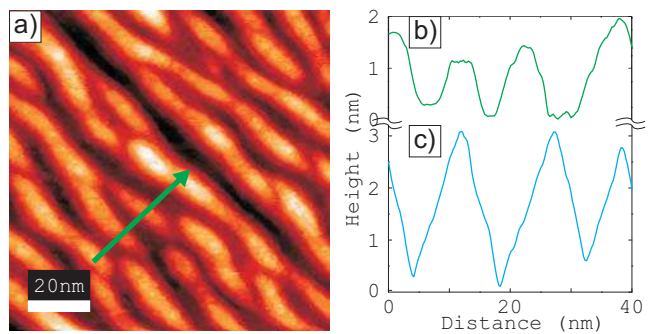}%
  \caption{\label{fig-fe-deposition}(a)~$\thicknm{100}$ STM image of wires Fe(\thickAL{2.5})/W
  (b)~Cross-section along the line shown in a (c)~Cross-section along the line shown in \figref{fig-template}c
  (same scales).}
  \end{center}
\end{figure}

The second step consists of the preparation of a non-magnetic template. It was inspired by reports
of the kinetic uniaxial roughening of films of the bcc elements Fe(110)\cite{bib-ALB93} and
W(110)\cite{bib-KOH00} during homoepitaxy at moderate temperature, explained by anisotropic
diffusion along the steps and the occurrence of an Ehrlich-Schwoebel barrier.
\subfigref{fig-template}a (resp.~c) shows the topography of a W(110) layer obtained upon
deposition at $\tempC{150}$ (resp. following nucleation at $\tempC{550}$ and subsequent growth at
$\tempC{150}$). A uniaxial array of trenches aligned along $[001]$ is thus obtained, with a period
of $\thicknm{4}$ (resp. $\thicknm{12}$). The STM cross-sections shown as insets reveal a depth of
$\thicknm{0.6-0.8}$ (resp. \thicknm{2-3}). The structure of the trenches can be probed using RHEED
with the electron beam along $[001]$\bracketsubfigref{fig-template}{b,d}. The diffraction patterns
consist of arrows whose half-angle reveals the slope of the microfacets. The arrows are much
sharper for trenches of larger depth because the coherence length along the facets is increased.
This allows us to determine accurately the angle of the facets to $\theta=\angledeg{18\pm1}$, in
perfect agreement with the triangular-like STM cross-sections of \figref{fig-fe-deposition}c. This
angle is unambiguously ascribed to facets of type $\{210\}$, with $\theta=\angledeg{18.43}$
expected. The monitoring of RHEED patterns during growth shows that $\angledeg{18}$ is a
stationary value, reached after a few nanometers of nominal thickness. The occurrence of a
stationary angle garanties that the shape of the trenches is not influenced by fluctuations of the
width of the trenches, and thus displays essentially no distribution. This is crucial as the
dispersion of MAE increases tremendously for nanosized systems in relation with the distribution
of structural environments\cite{bib-ROH06}. The existence of a stationary angle was postulated in
Ref.\cite{bib-KOH00}, however the analysis concluded to facets of type ${310}$ with
$\theta\sim\angledeg{26.57}$  for W. The discrepancy may come from the fact that the LEED patterns
used in Ref.\cite{bib-KOH00} were broad because trenches with a small period were studied. We
could produce templates with the same facets for Mo, Nb, V and Ta by deposition at room
temperature.

The third step consists of the fabrication of wires. Fe is deposited on the templates at
$\tempC{150}$\bracketfigref{fig-fe-deposition}, for which layer-by-layer growth occurs with PLD on
a smooth surface\cite{bib-FRU04,bib-FRU07}. For all deposits we have used W templates, which
remain perfectly stable up to $\tempC{300}$. For templates of small period the surface gradually
smoothes and becomes essentially flat for $\Theta=\thickAL{2}$(Atomic Layers; not shown here). The
details of the early stages of growth are easier to investigate for templates of large period. For
these STM cross-sections \bracketfigref{fig-fe-deposition}{b} reveal a flat level at the bottom of
the trenches, suggesting a progressive filling and thus yielding Fe wires with a triangular
cross-section. Disconnected wires can be formed up to $\Theta=\thickAL{2.5}$, beyond which
percolation sets in for the \thicknm{10} period.

We report below the magnetic properties of Fe wires prepared on W(110) templates with a mean
period of \unit[10]{nm} and an Fe nominal thickness of \unit[2.5]{AL}. The mean thickness and
width of the wires are \unit[1]{nm} and \unit[7]{nm}, respectively, as deduced from STM.

\begin{figure}
  \begin{center}
  \includegraphics[width=85.5mm]{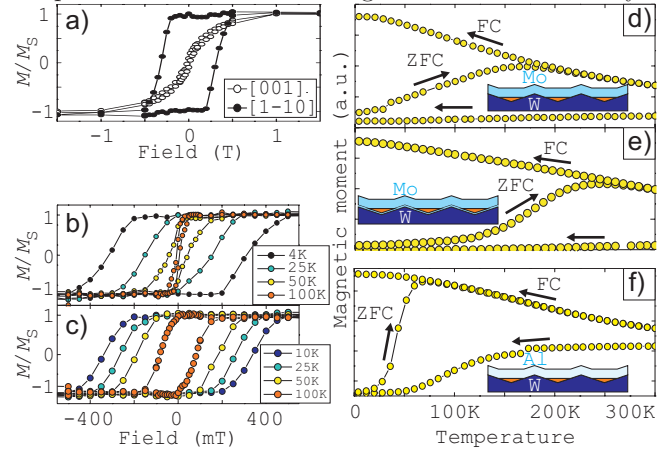}%
  \caption{\label{fig-squid}Magnetization curves of Fe(\thickAL{2.5}) wires with \thicknm{10} period.
  Mo/Fe/W wires (a)~at \tempK{10} in-plane along $[001]$ and $[1\overline{1}0]$ (b)~in-plane along $[001]$ at
   different temperatures; (c)~same as b, however for Mo/Fe/Mo/W wires, see text; (d-f)~In-plane field-cooled~(FC)
   and zero-field-cooled~(ZFC) magnetization of (d)~Mo/Fe/W (e)~Mo/Fe/Mo/W
  and (f)~Al/Fe/W wires. A schematic illustration of the samples is shown in insets.}
  \end{center}
\end{figure}

For Mo/Fe/W wires the easy axis of magnetization lies in-the-plane along the wires, \ie along
[001]\bracketsubfigref{fig-squid}a. The MAE is deduced from loops along [1$\overline{1}$0] like
$E_{\mathrm{a}}=\mu_{0}\int^{M_{S}}_{0}H \diff{M}\approx3\times\unit[10^{5}]{J/m^{3}}$ at
$\tempK{10}$. $E_{\mathrm{a}}$ originates from the sum of several contributions, among which only
the dipolar energy $E_{\mathrm{d}}$ can be estimated reliably. Other sources of MAE are surface
(N{\'e}el-type) expected to favor the [1$\overline{1}$0] direction for the (bottom) Fe/W
interface\cite{bib-GRA86} and [001] for the (top) Mo/Fe interface\cite{bib-FRU99c}, step-edge for
Fe/W, expected to favor [1$\overline{1}$0] \cite{bib-HAU98}; magneto-elastic -- unlike the case of
thin films, here the large density of steps is expected to induce a significant out-of-plane
mismatch and shear stress, so that no figure or even sign can be reliably predicted; finally the
bulk Fe MAE is negligible. Surprisingly $E_{\mathrm{a}}\approx E_{\mathrm{d}}$ for this sample
despite this complex situation, similarly to (Fe,Ag) self-organized arrays of
wires\cite{bib-FRU05d}.

The loops measured at different temperatures with $H//[001]$ have a rather square shape, and at
remanence full saturation is observed~\bracketsubfigref{fig-squid}b. The coercivity continuously
decreases with temperature while $\Ms$ remains essentially unchanged, suggesting a
superparamagnetic behavior. The ultimate blocking temperature determined by the "zero field
cooling/field cooling" (ZFC-FC) process is $\approx \unit[160]{K}$~\bracketsubfigref{fig-squid}d.
The \textsl{mean} value of $T_\mathrm{B}$ is around $\approx\unit[100]{K}$, defined as half-way up
the zero-field cooling remagnetization curve.

In the following we report on the use of surface MAE to tailor the magnitude of the MAE of the
wires and move towards features that would be required for applications like recording media, like
functionality at room temperature and adjustment of MAE. First, the very same wires were
fabricated after inserting a $\thickAL{1}$-thick underlayer of Mo on the W trenches. The magnetic
easy axis is again along the wires, however its magnitude is increased, which is consistent as
Fe/Mo interfaces favor alignement of magnetization along $[001]$\cite{bib-FRU99c}. This brings the
mean $T_\mathrm{B}$ from $\tempK{100}$ without underlayer to $\tempK{175}$ with
underlayer\bracketsubfigref{fig-squid}{e}. The coercive field is increased at all temperatures,
compare \subfigref{fig-squid}{b-c}. On the reverse $T_\mathrm{B}$ is lowered to $\tempK{40}$ for
Al/Fe/W\bracketsubfigref{fig-squid}{f}.

\section*{Conclusion}

We have demonstrated a novel route for the self-organization of arrays of planar nanowires on
nominally-flat commercial wafers. It is based on the fabrication of a template of trenches upon
the deposition under kinetic limitations of a non-magnetic material along a surface of uniaxial
symmetry, here W(110), followed by the filling of the bottom of the trenches by a magnetic
material under layer-by-layer deposition conditions, here Fe. Periods in the range
$\thicknm{4-12}$ were demonstrated. Owing to a self-limiting effect the micro-facets are all of
type ${210}$ with an angle of $\approx\angledeg{18\pm1}$, which is promising for attaining a low
distribution of physical properties. Concerning magnetism the easy axis of magnetization lies
in-the-plane along the wires. We have demonstrated the possibility to tailor its magnitude using
interface MAE with different capping- or underlayers, bringing the \textsl{mean} blocking
temperature to $\tempK{175}$ inserting an ultrathin underlayer of Mo.

B. B. and A. R. acknowledge financial support from French R{\'e}gion Rh{\^o}ne-Alpes (mobility program),
and FP6 EU-NSF program (STRP 016447 MagDot), respectively.

\section*{References}


\begin{thebibliography}{10}

\bibitem{bib-FRU03c}
P.~O. Jubert, J.~C. Toussaint, O. Fruchart, C. Meyer, and Y. Samson, \epl {\bf
  63},  135  (2003).

\bibitem{bib-GAM03b}
P. Gambardella, \jpcm {\bf 15},  S2533  (2003).

\bibitem{bib-GAM03}
P. Gambardella, S. Rusponi, M. Veronese, S.~S. Dhesi, C. Grazioli, A.
  Dallmeyer, I. Cabria, R. Zeller, P.~H. Dederichs, K. Kern, C. Carbone, and H.
  Brune, \science {\bf 300},  1130  (2003).

\bibitem{bib-REP00}
J. Repp, F. Moresco, G. Meyer, K.-H. Rieder, P. Hyldgaard, and M. Persson, \prl
  {\bf 85},  2981  (2000).

\bibitem{bib-WEI05}
N. Weiss, T. Cren, M. Epple, S. Rusponi, G. Baudot, S. Rohart, A. Tejeda, V.
  Repain, S. Rousset, P. Ohresser, F. Scheurer, P. Bencok, and H. Brune, \prl
  {\bf 95},  157204  (2005).

\bibitem{bib-MAS02}
{\em Semiconductor Quantum Dots}, {\em NanoScience and Technology}, edited by
  Y. Masumoto and T. Takagahara (Springer, Berlin, 2002).

\bibitem{bib-CRO05}
V. Cros, O. Boulle, J. Grollier, A. {Hamz{\'\i} c}, M. Mu{\~n}oz, L.~G.
  Pereira, and F. Petroff, \crp {\bf 6},  956  (2005).

\bibitem{bib-ALL05}
D.~A. Allwood, G. Xiong, C.~C. Faulkner, D. Atkinson, D. Petit, and R.~P.
  Cowburn, \science {\bf 309},  1688  (2005).

\bibitem{bib-PAR04}
S.~S.~P. Parkin, U.S. patents 6834005, 6898132, 6920062.

\bibitem{bib-HAU98}
J. Hauschild, U. Gradmann, and H.~J. Elmers, \apl {\bf 72},  3211  (1998).

\bibitem{bib-DAL00}
A. Dallmeyer, C. Carbone, W. Eberhardt, C. Pampuch, O. Rader, W. Gudat, P.
  Gambardella, and K. Kern, \prb {\bf 61},  R5133  (2000).

\bibitem{bib-FRU04}
O. Fruchart, M. Eleoui, J. Vogel, P.-O. Jubert, A. Locatelli, and A.
  Ballestrazzi, \apl {\bf 84},  1335  (2004).

\bibitem{bib-KER91}
K. Kern, H. Niehus, A. Schatz, P. Zeppenfeld, J. Goerge, and G. Comsa, \prl
  {\bf 67},  855  (1991).

\bibitem{bib-MOR03}
R. Moroni, D. Sekiba, F. {Buatier de Mongeot}, G. Gonella, C. Boragno, L.
  Mattera, and U.Valbusa, \prl {\bf 91},  167207/1  (2003).

\bibitem{bib-FRU07}
O. Fruchart, M. Eleoui, P.-O. Jubert, P. David, V. Santonacci, F. Cheynis, B.
  Borca, M. Hasegawa, and C. Meyer, \jpcm {\bf 19},    (2007), in press.

\bibitem{bib-FRU98b}
O. Fruchart, S. Jaren, and J. Rothman, \ass {\bf 135},  218  (1998).

\bibitem{bib-ALB93}
M. Albrecht, H. Fritzsche, and U. Gradmann, \ss {\bf 294},  1  (1993).

\bibitem{bib-KOH00}
U. K{\"o}hler, C. Jensen, C. Wolf, A.~C. Schindler, L. Brendel, and D. Wolf,
  \ss {\bf 454-456},  676  (2000).

\bibitem{bib-ROH06}
S. Rohart, V. Repain, A. Tejeda, P. Ohresser, F. Scheurer, P. Bencok, J.
  Ferr{\'e}, and S. Rousset, \prb {\bf 73},  165412  (2006).

\bibitem{bib-GRA86}
U. Gradmann, J. Korecki, and G. Waller, \apa {\bf A39},  101  (1986).

\bibitem{bib-FRU99c}
O. Fruchart, J.-P. Nozi\`eres, and D. Givord, \jmmm {\bf 207},  158  (1999).

\bibitem{bib-FRU05d}
B. Borca, O. Fruchart, and C. Meyer, \jap {\bf 99},  08Q514  (2005).

\end{thebibliography}

\end{document}